\documentclass[sigconf]{acmart}
\usepackage{multirow}
\usepackage{graphicx}
\usepackage{subcaption}
\AtBeginDocument{%
  }

\copyrightyear{2026}
\acmYear{2026}
\setcopyright{cc}
\setcctype{by}
\acmConference[MM '26] {Proceedings of the 34th ACM International Conference on Multimedia}{November 10--14, 2026}{Rio de Janeiro, Brazil}
\acmBooktitle{Proceedings of the 34th ACM International Conference on Multimedia (MM '26), November 10--14, 2026, Rio de Janeiro, Brazil}
\acmDOI{10.1145/3767308.3836213}
\acmISBN{979-8-4007-2213-4/2026/11}

\begin{document}

\title{Beyond Noisy Signals: Dual-Level Denoising for Multi-modal Sequential Recommendation}

\author{Jie Luo}
\orcid{0009-0003-5125-0560}
\affiliation{%
  \institution{University of Science and Technology of China}
  \city{Hefei}
  \country{China}
}
\email{luojie2000@mail.ustc.edu.cn}

\author{Qi Jin}
\orcid{0009-0002-5847-7779}
\affiliation{%
  \institution{University of Science and Technology of China}
  \city{Hefei}
  \country{China}
}
\email{jinqi200208@mail.ustc.edu.cn}

\author{Xinming Zhang}
\authornote{Corresponding authors.}
\orcid{0000-0002-8136-6834}
\affiliation{%
  \institution{University of Science and Technology of China}
  \city{Hefei}
  \country{China}
}
\email{xinming@ustc.edu.cn}

\renewcommand{\shortauthors}{Jie Luo, Qi Jin, and Xinming Zhang}

\begin{abstract}
Multi-modal Sequential Recommendation (SR) incorporates rich side information (e.g., textual and visual features) to enhance dynamic user preference modeling. However, existing frameworks inevitably suffer from a \textbf{Dual-Noise Dilemma}: (1) \textit{Feature-level redundancy} stemming from the semantic gap between generic pre-trained representations and fine-grained recommendation intent; and (2) \textit{Sequence-level stochasticity} induced by spurious interactions such as accidental clicks. To break this bottleneck, we propose \textbf{DDMSR}, a novel \textbf{D}ual-level \textbf{D}enoising \textbf{M}ulti-modal \textbf{S}equential \textbf{R}ecommendation framework that systematically purifies signals from both feature-topological and sequence-frequency perspectives. Specifically, we first design a graph-based feature denoising module that leverages Laplacian smoothing on item semantic graphs as a structural low-pass filter, effectively suppressing high-frequency semantic noise while preserving salient features. For sequence purification, we introduce a frequency-domain sequence denoising module, utilizing the Fast Fourier Transform and a learnable frequency filter to adaptively modulate the interaction spectrum and attenuate anomalous signals. Furthermore, a multi-modal contrastive alignment objective is incorporated to bridge the heterogeneity gap and enforce cross-modal semantic consistency. Extensive experiments on four public benchmark datasets demonstrate that DDMSR consistently outperforms state-of-the-art baselines, providing a highly robust and efficient solution for multi-modal sequential recommendation. The source code is available at: 
~\href{https://github.com/jluo00/DDMSR}{\textcolor{blue}{https://github.com/jluo00/DDMSR}}.
\end{abstract}




\maketitle

\section{INTRODUCTION}\label{Sec: intro}
Sequential Recommendation (SR) aims to model the evolutionary trajectory of users' historical interaction behaviors to capture their dynamic preferences and predict the next item of potential interest~\cite{sr_survey_1:conf/ijcai/wanghwcso19,sr_survey_2:journals/csur/wangcwsol22,sr_survey_3:journals/corr/abs-2412-12770}. With the continuous innovation of deep learning architectures, particularly the introduction of the Transformer~\cite{transformer:conf/nips/vaswanispujgkp17}, representative models such as SASRec~\cite{sasrec:conf/icdm/kangm18} and BERT4Rec~\cite{bert4rec:conf/cikm/sunlwploj19} have achieved significant success in capturing long- and short-term dependencies within sequences via self-attention mechanisms. However, the conventional SR paradigm relies heavily on item ID embeddings, which leads to severe cold-start issues for long-tail items and renders the models vulnerable to data sparsity in real-world scenarios~\cite{msr_survey:journals/csur/liuhxzgwlt25, morec:conf/sigir/yuanyslfypn23}. To break through these limitations, the Multi-modal Sequential Recommendation (Multi-modal SR) paradigm has emerged, aiming to enhance representation generalization by integrating side information such as textual descriptions and visual images. For instance, UniSRec~\cite{unisrec:conf/kdd/houmzldw22} utilizes a Mixture-of-Experts (MoE) network~\cite{moe:conf/iclr/shazeermmdlhd17} to learn transferable unified textual representations; MMSR~\cite{mmsr:conf/cikm/huglk23} leverages heterogeneous graph networks for adaptive multi-modal fusion; while TedRec~\cite{tedrec:conf/cikm/xutl0ww0cz24} and HM4SR~\cite{hm4sr:conf/www/zhang0s0025} enrich user modeling from the perspectives of frequency-domain integration and timestamp injection, respectively.

\begin{figure*}[t]
  \centering
   \includegraphics[width=0.9\linewidth]{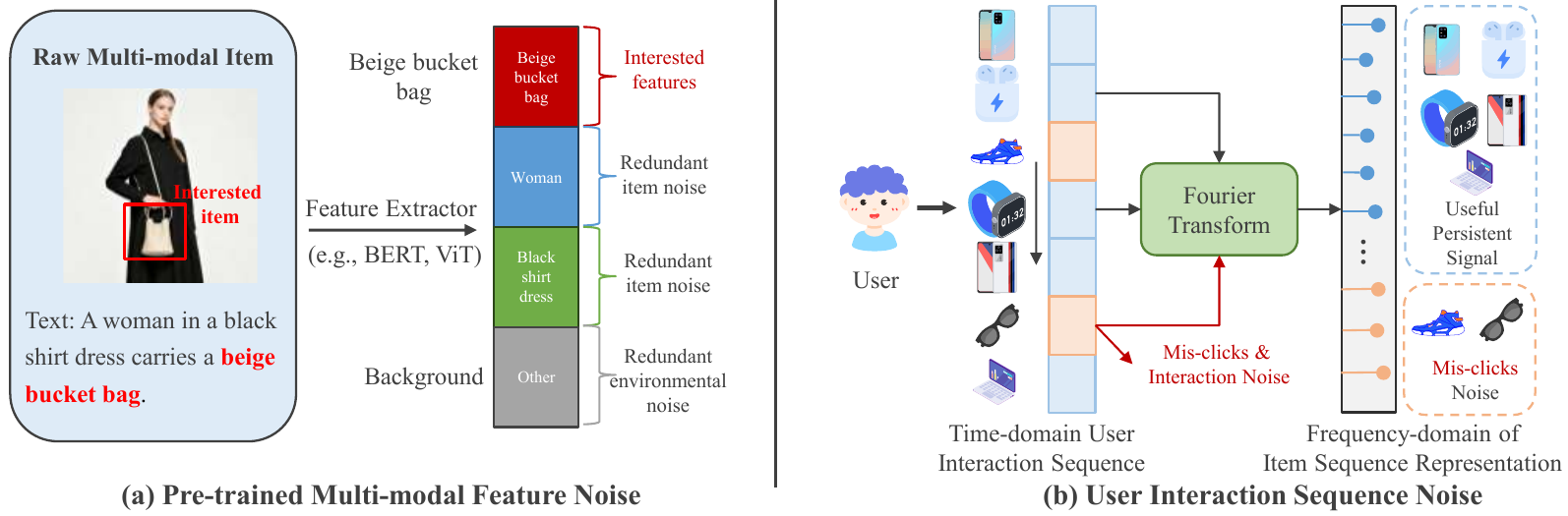}
   \caption{Illustration of multi-level noise in multi-modal recommendation. (a) Feature-level noise: Pre-trained encoders (e.g., ViT, BERT) often capture redundant, item-irrelevant details (e.g., background) misaligned with user intent. (b) Interaction-level noise: Raw sequences contain stochastic noise (e.g., mis-clicks). We transform sequences into the frequency domain to distinguish persistent interest (low-frequency) from abrupt noise (high-frequency). \textit{The image of the woman is AI-generated}.}
   \Description{}
   \label{fig1:toy example}
\end{figure*}

Despite the enriched item representations brought by multi-modal data, existing frameworks still encounter performance bottlenecks in practice. We observe that these models generally fall into a \textbf{``Dual-Noise Dilemma''}, which significantly restricts the upper bound of recommendation accuracy:
\begin{itemize}
    \item \textbf{Feature-level Noise}: The semantic gap between pre-trained representations and recommendation intent. Existing methods typically utilize dense vectors extracted from general-purpose pre-trained models (e.g., BERT~\cite{bert:conf/naacl/devlinclt19}, ViT~\cite{vit:conf/iclr/dosovitskiyb0wz21}) and project them into a unified embedding space via simple linear or non-linear mappings. However, these pre-trained models are fundamentally designed to capture coarse-grained, global semantics—such as the overall scene of an image or the general summary of a long text. This creates a non-negligible semantic gap between the pre-trained features and the fine-grained attributes that truly drive user preferences in recommendation scenarios (e.g., the specific silhouette of a garment or the unique texture of a product). As illustrated in Figure~\ref{fig1:toy example}(a), while a user's genuine interest may be focused on a specific ``beige bucket bag,'' the pre-trained encoders inevitably encode redundant background elements and irrelevant matching items into the feature space. Such feature-level noise subsequently contaminates the precision of user interest modeling.
    
    \item \textbf{Sequence-level Noise}: Spurious and stochastic interactions in the time domain. Prevailing models often operate under the idealized assumption that every interaction in a user's historical sequence faithfully reflects a stable and persistent preference. Nevertheless, real-world interaction sequences are inevitably perturbed by ``spurious interactions'' caused by accidental clicks, system exposure bias, or fleeting, impulsive demands. As depicted in Figure~\ref{fig1:toy example}(b), for a user whose core interest is electronics, the sporadic interactions with shoes or glasses within the sequence are likely high-frequency noise signals rather than indicators of long-term preference. Treating these noisy interactions as equivalent to authentic interests in the time domain will inevitably lead to biased estimations of the user's true intent, thereby degrading the overall recommendation quality.
\end{itemize}

To systematically address the aforementioned limitations, we propose \textbf{DDMSR}, a novel \textbf{D}ual-level \textbf{D}enoising \textbf{M}ulti-modal \textbf{S}equential \textbf{R}ecommendation framework that purifies recommendation signals from two complementary perspectives: feature-level denoising via graph topology and sequence-level denoising via frequency modulation. The framework consists of three key components.

\textit{Graph-based Feature Denoising.} To mitigate noise and redundancy in pre-trained multi-modal features, we construct sparse item-item semantic graphs from raw textual and visual representations, transcending the conventional isolated item paradigm. From a graph signal processing perspective, the neighborhood aggregation process inherently performs Laplacian smoothing on the graph manifold, acting as a low-pass filter that suppresses high-frequency noisy deviations inconsistent with local semantic neighborhoods while amplifying shared salient signals. Moreover, we design learnable adaptive residual connections to dynamically balance denoising and over-smoothing prevention, while implicitly performing feature imputation for long-tail items via semantic propagation.

\textit{Frequency-domain Sequence Denoising.} To address the inherent stochasticity and noise in interaction sequences, we introduce a frequency-domain adaptive modulation layer prior to the sequence encoder. Specifically, the multi-modal item embedding sequence is projected from the time domain into the frequency domain via the Fast Fourier Transform (FFT). A learnable frequency filter is then applied to perform global spectral modulation across the full frequency range. Rather than using a fixed low-pass filter, this data-driven design enables the model to adaptively suppress anomalous frequency components inconsistent with genuine user preferences—whether high-frequency spikes induced by accidental clicks or low-frequency drifts caused by transient interests—while selectively preserving and amplifying informative components with high predictive value. The modulated spectrum is then projected back to the time domain through the Inverse FFT, resulting in a robust and denoised sequence for preference modeling.

\textit{Multi-modal Contrastive Alignment.}
To bridge the heterogeneity gap between textual and visual feature spaces and facilitate coherent multi-modal fusion, we incorporate a contrastive learning objective that maximizes the mutual information between different modality representations of the same item, enforcing cross-modal semantic consistency.


Our main contributions are summarized as follows:
\begin{itemize}
    \item We identify and formulate the \textit{Dual-Noise Dilemma} in multi-modal sequential recommendation, highlighting the critical impact of feature redundancy and spurious interactions.
    \item We propose a novel dual-channel denoising framework \textbf{DDMSR} that integrates a graph-based spectral filter for feature refinement and an adaptive FFT-based modulator for sequence purification. Furthermore, a contrastive alignment objective is integrated to enforce cross-modal consistency.
    \item Extensive experiments on four real-world datasets demonstrate that our method significantly outperforms state-of-the-art baselines, and comprehensive ablation studies validate the efficacy of our dual-denoising design.
\end{itemize}

\section{RELATED WORKS}
In this section, we review related work on traditional ID-based sequential recommendation, multi-modal sequential recommendation and denoising sequential recommendation.

\subsection{Traditional Sequential Recommendation}
Sequential Recommendation (SR) aims to characterize the evolutionary patterns of user preferences from historical interaction trajectories to predict future behaviors~\cite{sr_survey_1:conf/ijcai/wanghwcso19,sr_survey_2:journals/csur/wangcwsol22}. Early paradigms primarily leveraged Markov Chains, such as FPMC~\cite{fpmc:conf/www/rendlefs10}, to capture low-order sequential transitions. However, these methods struggle to model the complex, long-range dependencies inherent in real-world sequences. 

The field has been revolutionized by deep learning architectures. RNN-based GRU4Rec~\cite{gru4rec:journals/corr/hidasikbt15} first introduced temporal modeling to SR, while CNN-based models like Caser~\cite{caser:conf/wsdm/tangw18} and NextItNet~\cite{nextitnet:conf/wsdm/yuankaj019} utilized convolutional filters to capture local patterns. Currently, Transformer-based models have established dominance due to their superior global dependency modeling. Specifically, SASRec~\cite{sasrec:conf/icdm/kangm18} employs unidirectional self-attention for generative modeling, whereas BERT4Rec~\cite{bert4rec:conf/cikm/sunlwploj19} adopts a bidirectional masked language modeling paradigm for richer contextual awareness.

Despite their efficacy, these ID-centric methods inherently treat items as isolated symbols, neglecting the rich multi-modal content. Such semantic isolation severely limits their representation generalization, leading to suboptimal performance in cold-start and sparse scenarios.

\subsection{Multi-modal Sequential Recommendation}
To transcend the limitations of ID-based modeling in cold-start and data-sparse scenarios, recent research has pivoted towards incorporating multi-modal side information. Early attempts like UniSRec~\cite{unisrec:conf/kdd/houmzldw22} utilize a Mixture-of-Experts (MoE) architecture to learn transferable item representations from text, while MMMLP~\cite{mmmlp:conf/www/liangzlzwl023} prioritizes computational efficiency by replacing complex attention mechanisms with a pure MLP-based backbone.
Subsequent studies have focused on more sophisticated fusion strategies. MISSRec~\cite{missrec:conf/mm/wangzwwllyzzx23} employs an interest-aware Transformer encoder-decoder with dynamic fusion to model sequence-level multi-modal interests.
MMSR~\cite{mmsr:conf/cikm/huglk23} disentangles sequential preferences into modality-independent and modality-synergistic patterns, employing an intent-oriented dynamic routing mechanism for adaptive fusion. More recently, researchers have explored specialized domains and architectures: TedRec~\cite{tedrec:conf/cikm/xutl0ww0cz24} enhances multi-modal modeling by performing sequence-level text-ID semantic fusion in the frequency domain; IISAN~\cite{iisan:conf/sigir/fug0kawj24} proposes a decoupled side-adaptation network to enable deep cross-modal interaction while preserving backbone parameters; and HM4SR~\cite{hm4sr:conf/www/zhang0s0025} incorporates time-aware MoE to fuse temporal dynamics into multi-modal modeling.

Although leveraging multi-modal features, existing frameworks still encounter performance bottlenecks due to an overlooked dual-noise dilemma: (1) feature-level redundancy inherent in generic pre-trained encoders, and (2) sequence-level stochasticity from accidental interactions. This combined noise impedes the precise characterization of true user preferences.

\subsection{Denoising Sequential Recommendation}
Denoising has emerged as a fundamental challenge in SR, with existing research bifurcating into two primary trajectories:
\textbf{(1) Interaction-level Denoising.} This line purifies behavior sequences from stochastic perturbations. Conventional methods filter noise either implicitly via frequency-domain modulation (e.g., FMLP-Rec~\cite{fmlp-rec:conf/www/zhouyzw22}, BSARec~\cite{bsarec:conf/aaai/shin0wp24}) or explicitly via structural inconsistency (e.g., HSD~\cite{hsd:conf/cikm/zhangdzhcl22}, SSDRec~\cite{ssdrec:conf/icde/zhangh00ts24}). Recently, generative and semantic paradigms have emerged: PDRec~\cite{pdrec:conf/aaai/maxmczlk24} and CaDiRec~\cite{cadirec:conf/cikm/cuiwhc024} leverage diffusion models for historical behavior reweighting and context-aware sequence augmentation, respectively, while IADSR~\cite{iadsr:conf/cikm/wu00z025} employs LLM-based alignment to filter noise via interest consistency.
\textbf{(2) Feature-level Denoising.} A parallel trajectory mitigates redundancy in pre-trained multi-modal representations. Pioneered by DiffuRec~\cite{diffurec:journals/tois/lisl24}'s distribution-based uncertainty modeling, recent works like DMMD4SR~\cite{dmmd4sr:conf/mm/luy25} and MDSBR~\cite{mdsbr:conf/recsys/liz25} explicitly repurpose diffusion processes to purify multi-modal features against domain-shift noise. Furthermore, broader multi-modal methods (e.g., DA-MRS~\cite{da-mrs:conf/kdd/xvlx0l0kl24}, DiffMM~\cite{diffmm:conf/mm/jiangx0llh24}) enhance feature purity by aligning content with user feedback or generating denoised semantic graphs.

Despite their progress, existing denoising paradigms suffer from two fundamental deficiencies: 
(1) Fragmented perspectives: Interaction-centric sequential models (e.g., FMLP-Rec~\cite{fmlp-rec:conf/www/zhouyzw22}, PDRec~\cite{pdrec:conf/aaai/maxmczlk24}) overlook item feature redundancy, while feature-centric approaches (e.g., MDSBR~\cite{mdsbr:conf/recsys/liz25}) often neglect sequence-level stochasticity. Notably, non-sequential models like DA-MRS~\cite{da-mrs:conf/kdd/xvlx0l0kl24} and DiffMM~\cite{diffmm:conf/mm/jiangx0llh24} fail to account for temporal interaction randomness, severely limiting their ability to capture dynamic preferences. 
(2) Computational inefficiency: Prevailing feature-denoising techniques rely heavily on iterative, computationally expensive generative processes (e.g., diffusion mechanisms in DMMD4SR~\cite{dmmd4sr:conf/mm/luy25}). These methods entail prohibitive overhead and fail to exploit the structural topology of items for principled filtering. 
In contrast, our framework provides a principled yet lightweight solution for synergistic purification. We move beyond costly generative modeling by leveraging Laplacian smoothing on item semantic graphs as a structural low-pass filter for features, while employing learnable frequency filters to adaptively modulate the interaction spectrum.

\section{METHODOLOGY}
\begin{figure*}[t]
  \centering
   \includegraphics[width=\linewidth]{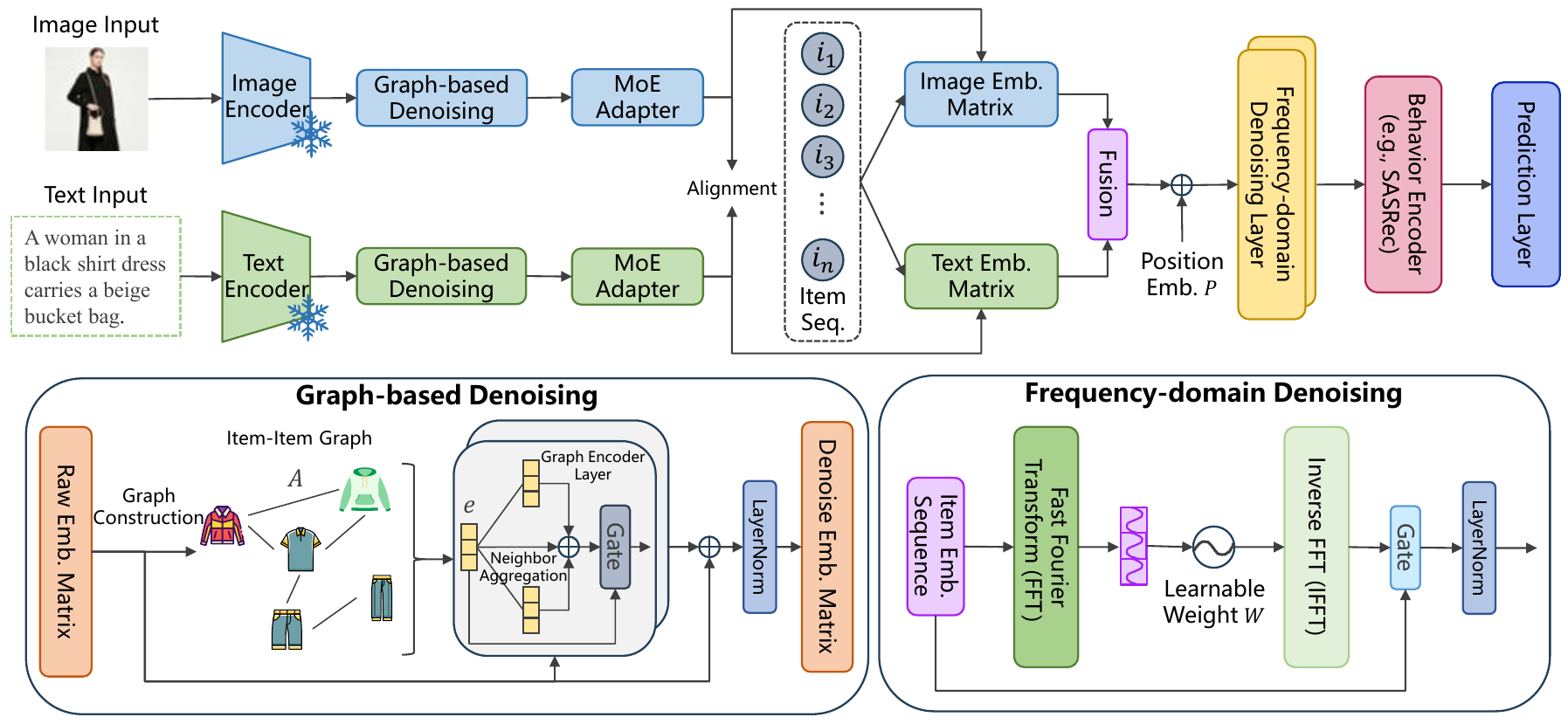}
   \caption{Overall framework of DDMSR.}
   \Description{}
   \label{fig2:overall framework}
\end{figure*}

In this section, we present the proposed \textbf{DDMSR} framework, as illustrated in Figure~\ref{fig2:overall framework}.

\subsection{Problem Formulation}
Let $\mathcal{U}$ and $\mathcal{I}$ denote the set of users and items, respectively. For each item $i \in \mathcal{I}$, it is associated with multi-modal information 
$\mathcal{M}_i = \{v_i, t_i\}$, where $v_i$ and $t_i$ represent the visual and textual features extracted by frozen pre-trained models, respectively.
For each user $u \in \mathcal{U}$, their historical interaction sequence is chronologically ordered as $S_u = [i_1, i_2, \dots, i_n]$, where $i_t \in \mathcal{I}$ 
denotes the $t$-th interacted item and $n = |S_u|$ is the sequence length. The goal of multi-modal sequential recommendation is to predict the next item $i_{n+1}$ 
that user $u$ is most likely to interact with, given $S_u$ and the associated multi-modal features:
\begin{equation}
    i_{n+1} = \arg\max_{i \in \mathcal{I}} 
    P(i_{n+1} = i \mid S_u, \{\mathcal{M}_{i_t}\}_{t=1}^{n}).
\end{equation}

\subsection{Multi-modal Feature Extraction and Denoising}

\subsubsection{Multi-modal Feature Extraction.}
To exploit the rich semantic information embedded in item multi-modal content, we employ frozen pre-trained encoders to extract raw multi-modal features. Specifically, a CLIP-ViT~\cite{clip:conf/icml/radfordkhrgasam21} is used for visual feature extraction, and RoBERTa~\cite{roberta:conf/acl/conneaukgcwggoz20} for textual feature extraction. For all items in $\mathcal{I}$, the raw visual and textual feature matrices are denoted as $\mathbf{F}_v \in \mathbb{R}^{|\mathcal{I}| \times d_v}$ and $\mathbf{F}_t \in \mathbb{R}^{|\mathcal{I}| \times d_t}$, respectively, where $d_v$ and $d_t$ are the pre-trained visual and text feature dimensions.

\subsubsection{Graph-based Multi-modal Feature Denoising.}
Since pre-trained encoders are optimized on large-scale general-domain corpora, the extracted features inevitably contain noise irrelevant to the recommendation task. To 
address this, we propose a graph-based denoising module grounded in graph signal processing theory: semantically similar items should have coherent representations in the feature space. By performing neighbor aggregation on an item semantic graph, we conduct Laplacian smoothing, which acts as a low-pass filter---suppressing high-frequency noisy deviations inconsistent with semantic neighbors while amplifying shared semantic signals across items.

\textbf{Item Semantic Graph Construction.}
For each modality $m \in \{v, t\}$, we compute pairwise cosine similarities between raw item features to construct a similarity matrix $\mathbf{S}^m$, where the $(a,b)$-th entry is defined as:
\begin{equation}
    s^m_{a,b} = \frac{(\mathbf{e}^m_a)^\top \mathbf{e}^m_b}
    {\|\mathbf{e}^m_a\| \cdot \|\mathbf{e}^m_b\|},
\end{equation}
where $\mathbf{e}^m_a$ and $\mathbf{e}^m_b$ are the modality-$m$ feature vectors of items $a$ and $b$, respectively. To prevent over-smoothing caused by dense connectivity and reduce computational overhead, we retain only the top-$K$ most similar neighbors for each item, yielding a sparse adjacency matrix $\mathbf{A}^m$. The adjacency matrix is then symmetrically normalized as:
\begin{equation}
    \hat{\mathbf{A}}^m = (\mathbf{D}^m)^{-1/2} 
    \mathbf{A}^m (\mathbf{D}^m)^{-1/2},
\end{equation}
where $\mathbf{D}^m$ is the corresponding degree matrix.

\textbf{Adaptive Graph-based Feature Denoising.}
Based on the constructed semantic graph, we perform $L_g$ layers of graph convolutional aggregation. At the $l$-th layer, item features are updated via an adaptive weighted 
combination of self features and neighbor-aggregated features:
\begin{equation}
    \mathbf{h}^{(l)} = w^{(l)} \cdot \mathbf{h}^{(l-1)} 
    + (1 - w^{(l)}) \cdot \hat{\mathbf{A}}^m 
    \mathbf{h}^{(l-1)},
\end{equation}
where $w^{(l)} = \sigma(\alpha^{(l)})$ is a learnable adaptive weight at layer $l$, $\sigma(\cdot)$ is the Sigmoid function, and $\mathbf{h}^{(0)} = \mathbf{F}_m$. This learnable weight enables the model to automatically balance the contribution of self features and aggregated neighbor features: items with high-quality features tend to retain more self information, while long-tail items with sparse signals benefit more from neighboring high-quality features. Finally, a residual connection and layer normalization are applied to obtain the denoised feature matrix:
\begin{equation}
    \tilde{\mathbf{F}}_m = \text{LayerNorm}
    (\mathbf{h}^{(L_g)} + \mathbf{F}_m).
\end{equation}

\subsubsection{MoE Adapter.}
To effectively project the denoised multi-modal features into a unified semantic space and facilitate cross-modal alignment, we employ Mixture-of-Experts (MoE) adapters 
to map $\tilde{\mathbf{F}}_v$ and $\tilde{\mathbf{F}}_t$ into unified representations $\mathbf{E}_v, \mathbf{E}_t \in \mathbb{R}^{|\mathcal{I}| \times d}$. Each MoE layer 
comprises $K$ expert networks, and the output for input $x$ is computed as a weighted sum of expert outputs:
\begin{equation}
    \tilde{x} = \sum_{k=1}^{K} g_k \cdot 
    \text{Expert}_k(x),
\end{equation}
where each expert is instantiated as a two-layer MLP with Dropout and ReLU activation~\cite{relu:journals/jmlr/glorotbb11}:
\begin{equation}
    \text{Expert}_k(x) = \phi(\mathbf{W}_2 \cdot 
    \text{Dropout}(\phi(\mathbf{W}_1 \cdot 
    \text{Dropout}(x) + \mathbf{b}_1)) + \mathbf{b}_2).
\end{equation}
Following UniSRec~\cite{unisrec:conf/kdd/houmzldw22}, we adopt a noisy gating mechanism to encourage diverse expert utilization and improve load balancing:
\begin{equation}
    g = \text{Softmax}(x\mathbf{W}_g + \epsilon \odot 
    \text{Softplus}(x\mathbf{W}_{noise})), \quad 
    \epsilon \sim \mathcal{N}(0, 1),
\end{equation}
where $\mathbf{W}_g$ and $\mathbf{W}_{noise}$ are learnable routing and noise projection matrices, respectively.

\subsection{Sequence Modeling with Frequency-domain Denoising}

\subsubsection{Multi-modal Sequence Construction.}
Given the denoised and MoE-adapted multi-modal feature matrices $\mathbf{E}_v, \mathbf{E}_t \in \mathbb{R}^{|\mathcal{I}| \times d}$, for user interaction 
sequence $S_u = [i_1, i_2, \dots, i_n]$, we retrieve the corresponding visual and textual feature sequences by index lookup and fuse them into a unified item embedding 
sequence:
\begin{equation}\label{eq:fusion}
    \mathbf{H} = \text{Dropout}\left(\text{LayerNorm}\left(
    \text{Fusion}(\mathbf{E}_v[S_u], \mathbf{E}_t[S_u]) 
    + \mathbf{P}\right)\right),
\end{equation}
where $\mathbf{P} \in \mathbb{R}^{n \times d}$ is a learnable positional embedding, and $\text{Fusion}(\cdot)$ can be instantiated as element-wise addition, concatenation, or attention-based fusion.

\subsubsection{Frequency-domain Sequence Denoising.}
User interaction sequences inevitably contain noisy interactions (\emph{e.g.}, accidental clicks) that do not reflect genuine user preferences. Feeding such noisy 
sequences directly into the sequential encoder would impair user preference modeling. Inspired by prior works~\cite{fmlp-rec:conf/www/zhouyzw22, bsarec:conf/aaai/shin0wp24, tasif:conf/wsdm/luozz026}, we design a frequency-domain denoising module consisting of $L_f$ stacked filtering layers to address this, which is inserted prior to the sequential encoder. The underlying intuition is that noisy interactions tend to manifest as high-frequency components in the frequency domain, while stable user preferences correspond to persistent low-frequency signals. By modulating the frequency spectrum with learnable filters, the module adaptively suppresses noise signals while preserving informative sequential patterns.

Formally, given the item embedding sequence $\mathbf{H} \in \mathbb{R}^{n \times d}$, each frequency-domain denoising layer proceeds as follows. First, the sequence is transformed from the time domain to the frequency domain via the Fast Fourier Transform (FFT):
\begin{equation}
    \mathbf{H}_f = \text{FFT}(\mathbf{H}),
\end{equation}
where $\mathbf{H}_f \in \mathbb{C}^{n \times d}$ is the complex spectrum tensor. A learnable complex-valued filter $\mathbf{W} \in \mathbb{C}^{n \times d}$ is then applied to modulate the 
spectrum via element-wise multiplication:
\begin{equation}
    \tilde{\mathbf{H}}_f = \mathbf{W} \odot 
    \mathbf{H}_f.
\end{equation}
The modulated spectrum is then transformed back to the time domain via the Inverse Fast Fourier Transform (IFFT):
\begin{equation}
    \tilde{\mathbf{H}} = \text{IFFT}
    (\tilde{\mathbf{H}}_f).
\end{equation}
To adaptively balance the contributions of the frequency-filtered features and the original features, we introduce a gating network for adaptive fusion, replacing the fixed residual connection:
\begin{equation}
    \alpha = \sigma\!\left(\text{MLP}([\mathbf{H} 
    \,\|\, \tilde{\mathbf{H}}])\right),
\end{equation}
\begin{equation}
    \bar{\mathbf{H}} = \text{LayerNorm}\!\left(
    \alpha \odot \tilde{\mathbf{H}} + 
    (1 - \alpha) \odot \mathbf{H}\right),
\end{equation}
where $\alpha \in (0,1)$ is a position-wise gating weight and $\sigma(\cdot)$ denotes the Sigmoid function. This design enables the model to adaptively regulate the denoising intensity at each sequence position, effectively suppressing noise while retaining valuable sequential patterns. After $L_f$ layers of frequency-domain denoising, 
the refined sequence representation is fed into a sequential encoder (\emph{e.g.}, SASRec) for user preference modeling.

\subsubsection{Sequential Modeling.}
After frequency-domain denoising, the refined sequence representation $\bar{\mathbf{H}} \in \mathbb{R}^{n \times d}$ is fed into a Transformer-based sequential encoder to 
capture item transition patterns and model the evolution of user preferences. We stack $L_t$ Transformer layers, each consisting of a Multi-Head Self-Attention (MHSA) 
sub-layer and a point-wise Feed-Forward Network (FFN), with residual connections and layer normalization applied at each sub-layer:
\begin{equation}
\begin{aligned}
    \mathbf{S}^{(l)} &= \text{MHSA}(\mathbf{H}^{(l-1)}) 
    + \mathbf{H}^{(l-1)}, \\
    \mathbf{H}^{(l)} &= \text{FFN}(\text{LayerNorm}(
    \mathbf{S}^{(l)})) + \mathbf{S}^{(l)},
\end{aligned}
\end{equation}
where $1 \le l \le L_t$ and $\mathbf{H}^{(0)} = \bar{\mathbf{H}}$. The MHSA sub-layer captures long-range dependencies between items at different positions, while the FFN introduces non-linearity for richer representation learning. Finally, we extract the hidden state at the last valid position $t$ from the output of the last Transformer layer $\mathbf{H}^{(L_t)}$ as the user preference representation:
\begin{equation}
    \mathbf{s}_u = \mathbf{H}^{(L_t)}[t].
\end{equation}

\subsection{Model Training}
We adopt a multi-task learning framework that jointly optimizes the primary recommendation task and an auxiliary cross-modal contrastive learning objective.

\noindent\textbf{Recommendation Loss.}
Given the user representation $\mathbf{s}_u \in \mathbb{R}^{1 \times d}$ produced by the sequential encoder, we compute matching scores via dot product with item embeddings, scaled by a temperature parameter $\tau_{rec}$. The primary recommendation task is optimized with a cross-entropy loss:
\begin{equation}
    \mathcal{L}_{rec} = -\frac{1}{|\mathcal{B}|}
    \sum_{u \in \mathcal{B}} \log 
    \frac{\exp(\mathbf{s}_u^\top \mathbf{e}_{y_u} 
    / \tau_{rec})}{\sum_{i \in \mathcal{I}} 
    \exp(\mathbf{s}_u^\top \mathbf{e}_i / \tau_{rec})},
\end{equation}
where $\mathcal{B}$ denotes the training batch, $\mathbf{e}_{y_u} \in \mathbb{R}^{1 \times d}$ is the embedding of the ground-truth next item for user $u$, and $\mathcal{I}$ is the full item set. The item embedding is defined as $\mathbf{e}_i = \mathbf{e}_i^t + \mathbf{e}_i^v$, where $\mathbf{e}_i^t \in \mathbf{E}_t$ and $\mathbf{e}_i^v \in \mathbf{E}_v$.

\noindent\textbf{Cross-modal Contrastive Learning Loss.}
To bridge the heterogeneity gap between feature spaces of different modalities, we introduce an auxiliary cross-modal contrastive learning objective. For each item $i$ in the batch, its textual and visual embeddings are treated as a positive pair, while embeddings of other items in the batch serve as negatives. We adopt the InfoNCE~\cite{infonce:journals/corr/abs-1807-03748} loss for cross-modal semantic alignment:
\begin{equation}
    \mathcal{L}_{cl} = -\frac{1}{|\mathcal{B}|}
    \sum_{i \in \mathcal{B}} \left[
    \log \frac{\exp(\mathbf{e}_i^{t\top} \mathbf{e}_i^v 
    / \tau_{cl})}{\sum_{j \in \mathcal{B}} 
    \exp(\mathbf{e}_i^{t\top} \mathbf{e}_j^v / \tau_{cl})}
    + \log \frac{\exp(\mathbf{e}_i^{v\top} \mathbf{e}_i^t 
    / \tau_{cl})}{\sum_{j \in \mathcal{B}} 
    \exp(\mathbf{e}_i^{v\top} \mathbf{e}_j^t / \tau_{cl})}
    \right],
\end{equation}
where $\tau_{cl}$ is the contrastive temperature parameter. By maximizing the mutual information between the two modality representations of the same item, this objective effectively reduces semantic misalignment and enhances the robustness of item representations.

\noindent\textbf{Overall Objective.}
The final training objective is a weighted combination of the recommendation loss and the contrastive loss:
\begin{equation}
    \mathcal{L} = \mathcal{L}_{rec} + 
    \lambda \mathcal{L}_{cl},
\end{equation}
where $\lambda$ is a hyperparameter that controls the relative contribution of the contrastive objective.

\subsection{Prediction}
Given the user representation $\mathbf{s}_u$ produced by the sequential encoder, we compute matching scores for all candidate items via the dot product with their multi-modal representations:
\begin{equation}
    \hat{\mathbf{y}}_u = \mathbf{s}_u (\mathbf{E}_t + \mathbf{E}_v)^\top,
\end{equation}
where $\mathbf{E}_t$ and $\mathbf{E}_v$ denote the denoised and MoE-adapted textual and visual embedding matrices, respectively. $\hat{\mathbf{y}}_u \in \mathbb{R}^{|\mathcal{I}|}$ represents the predicted scores over all items. The items are ranked according to these scores, and the top-$K$ items are selected for recommendation.

\section{EXPERIMENT}
In this section, we conduct extensive experiments to verify the effectiveness of our DDMSR. 

\begin{table}[t]
    \centering
    \caption{Statistics of the datasets after preprocessing.}
    \label{tab1:dataset_statistics}
    \begin{tabular}{lcccc}
        \toprule
        Dataset& Beauty & Sports & Toys & MicroLens  \\
        \midrule
        \# Users & 22,363 & 35,598 & 19,412 & 98,129 \\
        \# Items & 12,101 & 18,357 & 11,924 & 17,228 \\
        \# Avg. Actions / User & 7.9 & 7.3 & 7.6 & 6.2 \\
        \# Avg. Actions / Item & 14.6 & 14.2 & 12.4 & 35.3 \\
        \# Actions & 176,139 & 260,739 & 148,185 & 607,045 \\
        Sparsity & 99.93\% & 99.96\% & 99.94\% & 99.96\% \\
        \bottomrule
    \end{tabular}
\end{table}
\begin{table*}[t]
\centering
\setlength{\tabcolsep}{1mm}
\caption{Performance comparison of DDMSR and baselines. 
\textbf{Bold} and \underline{underline} denote the best and second-best results, respectively. 
\textit{Impr.} denotes the improvement over the best baseline. 
* indicates that the improvement over the best baseline is statistically significant according to paired t-tests ($p$ < 0.01). 
R@K and N@K represent Recall@K and NDCG@K, respectively.}
\label{tab:overall}
\resizebox{\textwidth}{!}{
    \begin{tabular}{clcccccccccccc}
    \toprule
    \multirow{2}{*}{Dataset} 
    & \multirow{2}{*}{Metric} 
    & \multicolumn{4}{c}{ID-based} 
    & \multicolumn{6}{c}{Multi-modal} 
    & \multirow{2}{*}{Ours} 
    & \multirow{2}{*}{\textit{Impr.}} \\
    \cmidrule(lr){3-6} \cmidrule(lr){7-12}
    & 
    & SASRec & BERT4Rec & FEARec & BSARec 
    & UniSRec & MISSRec & MP4SR & TedRec & HM4SR & DMMD4SR
    & 
    & \\
    \midrule
    
    \multirow{4}{*}{Beauty}
    & R@10 & 0.0842 & 0.0402 & 0.0855 & 0.0869 & 0.0892 & \underline{0.0978} & 0.0745 & 0.0879 & 0.0776 & 0.0828 & \textbf{0.1056*} & +7.98\% \\
    & R@20 & 0.1192 & 0.0678 & 0.1206 & 0.1226 & 0.1262 & \underline{0.1417} & 0.1067 & 0.1224 & 0.1068 & 0.1171 & \textbf{0.1564*} & +10.37\% \\
    & N@10 & 0.0416 & 0.0173 & 0.0422 & 0.0431 & 0.0441 & 0.0488 & 0.0400 & \underline{0.0512} & 0.0485 & 0.0413 & \textbf{0.0535*} & +4.49\% \\
    & N@20 & 0.0504 & 0.0243 & 0.0509 & 0.0522 & 0.0535 & 0.0597 & 0.0479 & \underline{0.0599} & 0.0559 & 0.0499 & \textbf{0.0663*} & +10.68\% \\
    
    \midrule
    \multirow{4}{*}{Sports}
    & R@10 & 0.0480 & 0.0387 & 0.0497 & 0.0485 & 0.0534 & \underline{0.0555} & 0.0332 & 0.0519 & 0.0446 & 0.0454 & \textbf{0.0643*} & +15.86\% \\
    & R@20 & 0.0703 & 0.0606 & 0.0725 & 0.0716 & 0.0792 & \underline{0.0812} & 0.0511 & 0.0752 & 0.0623 & 0.0662 & \textbf{0.0969*} & +19.33\% \\
    & N@10 & 0.0225 & 0.0163 & 0.0232 & 0.0228 & 0.0247 & 0.0266 & 0.0167 & \underline{0.0288} & 0.0270 & 0.0215 & \textbf{0.0332*} & +15.28\% \\
    & N@20 & 0.0282 & 0.0218 & 0.0288 & 0.0286 & 0.0312 & 0.0326 & 0.0212 & \underline{0.0348} & 0.0315 & 0.0267 & \textbf{0.0413*} & +18.68\% \\
    
    \midrule
    \multirow{4}{*}{Toys}
    & R@10 & 0.0912 & 0.0473 & 0.0920 & 0.0952 & 0.0954 & \underline{0.1095} & 0.0783 & 0.0900 & 0.0889 & 0.0893 & \textbf{0.1153*} & +5.30\% \\
    & R@20 & 0.1240 & 0.0706 & 0.1257 & 0.1314 & 0.1328 & \underline{0.1546} & 0.1084 & 0.1230 & 0.1179 & 0.1204 & \textbf{0.1657*} & +7.18\% \\
    & N@10 & 0.0445 & 0.0290 & 0.0450 & 0.0464 & 0.0452 & 0.0526 & 0.0447 & 0.0525 & \underline{0.0552} & 0.0439 & \textbf{0.0563*} & +1.99\% \\
    & N@20 & 0.0527 & 0.0348 & 0.0536 & 0.0555 & 0.0546 & \underline{0.0637} & 0.0523 & 0.0608 & 0.0629 & 0.0519 & \textbf{0.0690*} & +8.32\% \\

    \midrule
    \multirow{4}{*}{MicroLens}
    & R@10 & 0.0839 & 0.0454 & 0.0866 & 0.0912 & 0.0932 & 0.0837 & 0.0862 & \underline{0.0940} & 0.0824 & 0.0836 & \textbf{0.1019*} & +8.40\% \\
    & R@20 & 0.1234 & 0.0713 & 0.1246 & 0.1342 & \underline{0.1378} & 0.1273 & 0.1285 & 0.1358 & 0.1239 & 0.1229 & \textbf{0.1504*} & +9.14\% \\
    & N@10 & 0.0400 & 0.0231 & 0.0418 & 0.0444 & 0.0456 & 0.0414 & 0.0429 & \underline{0.0512} & 0.0438 & 0.0405 & \textbf{0.0520*} & +1.56\% \\
    & N@20 & 0.0499 & 0.0296 & 0.0513 & 0.0552 & 0.0567 & 0.0520 & 0.0536 & \underline{0.0618} & 0.0542 & 0.0504 & \textbf{0.0642*} & +3.88\% \\
    
    \bottomrule
    \end{tabular}
}
\end{table*}

\subsection{Experimental Settings}

\subsubsection{Datasets}
We evaluate our method on four widely used benchmarks spanning e-commerce and short-video platforms.

\begin{itemize}
    \item \textbf{Amazon\footnote{\url{http://jmccauley.ucsd.edu/data/amazon/}}~\cite{Amazon:conf/sigir/McAuleyTSH15}}: We select three sub-categories from this large-scale e-commerce dataset: \textit{Beauty}, \textit{Sports and Outdoors}, and \textit{Toys and Games} (hereafter \textit{Beauty}, \textit{Sports}, and \textit{Toys}). Textual features are constructed by concatenating the product \texttt{title}, \texttt{brand}, \texttt{categories}, and \texttt{description}. Product images serve as the visual modality.

    \item \textbf{MicroLens-100k\footnote{\url{https://github.com/westlake-repl/MicroLens/}}~\cite{MicroLens-100k:journals/corr/abs-2309-15379}:} This is a large-scale multi-modal dataset collected from a popular short-video platform, containing interaction data from 100,000 users. We specifically utilize the video cover thumbnails (covers) as visual features. For textual information, we combine the video \texttt{title} and associated \texttt{tags} to construct the text inputs.
\end{itemize}

Following Hou~\emph{et al.}~\cite{unisrec:conf/kdd/houmzldw22}, we filter out users and items with fewer than five interactions, and adopt the leave-one-out strategy to partition each sequence into training, validation, and test sets. Dataset statistics are summarized in Table~\ref{tab1:dataset_statistics}. 
Textual and visual features are extracted using RoBERTa~\cite{roberta:conf/acl/conneaukgcwggoz20} and CLIP ViT-B/32~\cite{clip:conf/icml/radfordkhrgasam21}, respectively.

\subsubsection{Evaluation Metrics}
We adopt Recall@$K$ and NDCG@$K$ ($K \in \{10, 20\}$) as evaluation metrics. Recall@$K$ measures the proportion of relevant items retrieved within the top-$K$ list, while NDCG@$K$ further accounts for their ranking positions. To avoid sampling bias, we 
conduct full-ranking evaluation~\cite{Metrics:conf/kdd/KricheneR20} over the entire item set.

\subsubsection{Baseline Models}
We compare our method against two categories of baselines. \textbf{ID-based models} include SASRec~\cite{sasrec:conf/icdm/kangm18}, 
BERT4Rec~\cite{bert4rec:conf/cikm/sunlwploj19}, 
FEARec~\cite{fearec:conf/sigir/duyzqz0ls23}, and 
BSARec~\cite{bsarec:conf/aaai/shin0wp24}. 
\textbf{Multi-modal models} include UniSRec~\cite{unisrec:conf/kdd/houmzldw22}, 
MISSRec~\cite{missrec:conf/mm/wangzwwllyzzx23}, 
MP4SR~\cite{mp4sr:journals/tors/zhangzzs25}, 
TedRec~\cite{tedrec:conf/cikm/xutl0ww0cz24}, 
HM4SR~\cite{hm4sr:conf/www/zhang0s0025}, and 
DMMD4SR~\cite{dmmd4sr:conf/mm/luy25}. To ensure a fair comparison, UniSRec, MISSRec and MP4SR are trained from scratch without pre-training on external datasets. More details can be found in the Appendix.

\subsubsection{Implementation Details}
All models are implemented within the open-source RecBole framework~\cite{RecBole:conf/cikm/ZhaoMHLCPLLWTMF21} to ensure fair and reproducible evaluation. All models are optimized using the Adam optimizer~\cite{Adam:journals/corr/KingmaB14} with a learning rate of 0.001. Hyperparameters of baseline models are tuned following the search spaces specified 
in their respective original papers. For our method, the number of top-$K$ neighbors $K$ for item semantic graph construction are searched in $\{5, 10, 15, 20\}$, and the number of frequency-domain denoising layers $L_f$ is searched in $\{1, 2, 3\}$. The number of graph convolutional layers $L_g$ is fixed to $2$, and the fusion function $\text{Fusion}(\cdot)$ in Eq.~(\ref{eq:fusion}) is element-wise addition. The contrastive loss weight $\lambda$ is set to $0.1$ based on preliminary experiments. All other hyperparameters follow the settings of UniSRec~\cite{unisrec:conf/kdd/houmzldw22}. All final results are averaged over five independent runs with different random seeds. All experiments are conducted on a NVIDIA GeForce RTX 4090 GPU (24 GB).

\subsection{Overall Performance Comparison}
Table~\ref{tab:overall} reports the overall performance of all models on four datasets, from which we draw the following observations.

Compared to classical SASRec and BERT4Rec, FEARec and BSARec incorporate frequency-domain perspectives to capture periodic signals and suppress sequence noise, yielding consistent improvements across datasets and validating the effectiveness of frequency-domain analysis for sequential recommendation. Multi-modal models generally outperform ID-based methods, demonstrating that multi-modal content features effectively alleviate data sparsity and cold-start issues. Among them, MISSRec achieves strong performance 
by designing an interest-aware encoder-decoder architecture that captures sequence-level multi-modal user interests with a dynamic fusion module, while TedRec enhances multi-modal modeling by performing sequence-level text-ID semantic fusion in the frequency domain. The two methods alternately attain the best baseline results across most datasets, jointly underscoring the importance of multi-modal representation learning and sequence-level semantic fusion. However, DMMD4SR---despite its diffusion-based denoising design---underperforms other multi-modal baselines across all four datasets, suggesting that the complex forward-reverse diffusion process is difficult to optimize in recommendation settings and its denoising objective is not well aligned with the recommendation task.

Our method DDMSR consistently achieves state-of-the-art performance, with improvements of up to +19.33\% on Recall@20 (Sports) and +10.68\% on NDCG@20 (Beauty) over the strongest baseline. This stems from three complementary designs: (1) graph-based denoising suppresses high-frequency noise in pre-trained features via Laplacian smoothing while compensating for long-tail items through semantic neighbor aggregation; (2) frequency-domain denoising adaptively filters noisy interaction signals in user sequences, preserving stable low-frequency preference components; and (3) the cross-modal contrastive objective bridges the heterogeneity gap between modality feature spaces for coherent multi-modal fusion.

\begin{table}[!t]
    \caption{Ablation study of our DDMSR (Recall@20).}
    \label{tab4:ablation}
    \begin{tabular}{l|cccc}
        \hline
        Variant & Beauty & Sports & Toys & MicroLens \\
        \hline
        w/o GFD & 0.1502 & 0.0935 & 0.1567 & 0.1425 \\
        w/o FSD & 0.1524 & 0.0940 & 0.1638 & 0.1461 \\
        w/o CL & 0.1505 & 0.0949 & 0.1610 & 0.1484 \\
        w/o Text & 0.1488 & 0.0846 & 0.1539 & 0.1431 \\
        w/o Image & 0.1471 & 0.0900 & 0.1576 & 0.1401 \\
        \hline
        DDMSR & \textbf{0.1564} & \textbf{0.0969} & \textbf{0.1657} & \textbf{0.1504} \\
        \hline
    \end{tabular}
\end{table}
\subsection{Ablation Study}

To assess the contribution of each component, we design five model variants and report results in Table~\ref{tab4:ablation}.

\begin{itemize}

    \item \textbf{w/o GFD} removes the \textbf{Graph-based Feature Denoising} (GFD) module, directly feeding raw pre-trained features into the MoE adapter.
    
    \item \textbf{w/o FSD} removes the \textbf{Frequency-domain Sequence Denoising} (FSD) layers before the sequential encoder, directly feeding the fused item embeddings to the Transformer.
    
    \item \textbf{w/o CL} removes the cross-modal contrastive learning (CL) loss $\mathcal{L}_{cl}$ and optimizes only the recommendation loss $\mathcal{L}_{rec}$.
    
    \item \textbf{w/o Text} removes the textual modality, performing sequence modeling using visual features only.
    
    \item \textbf{w/o Image} removes the visual modality, performing sequence modeling using textual features only.

\end{itemize}

The results lead to the following observations: \textbf{(1)} Removing \textbf{w/o GFD} causes a consistent performance drop across all datasets, confirming that the graph-based denoising module effectively suppresses high-frequency redundant noise in pre-trained features and is critical for improving item representation quality; \textbf{(2)} The degradation of \textbf{w/o FSD} validates the importance of frequency-domain denoising in filtering noisy interactions from user behavior sequences and preserving stable low-frequency user preference signals; \textbf{(3)} The performance decline of \textbf{w/o CL} demonstrates that the cross-modal contrastive objective plays a non-trivial role in bridging the heterogeneity gap between textual and visual feature spaces, enhancing the semantic consistency of multi-modal representations; \textbf{(4)} Both \textbf{w/o Text} and \textbf{w/o Image} lead to notable performance degradation, confirming that textual and visual modalities carry complementary information that cannot be fully captured by either modality alone.
Overall, the full model consistently achieves the best performance among all variants, justifying the design necessity of each component. 

\begin{figure}[t]
  \centering
   \includegraphics[width=\linewidth]{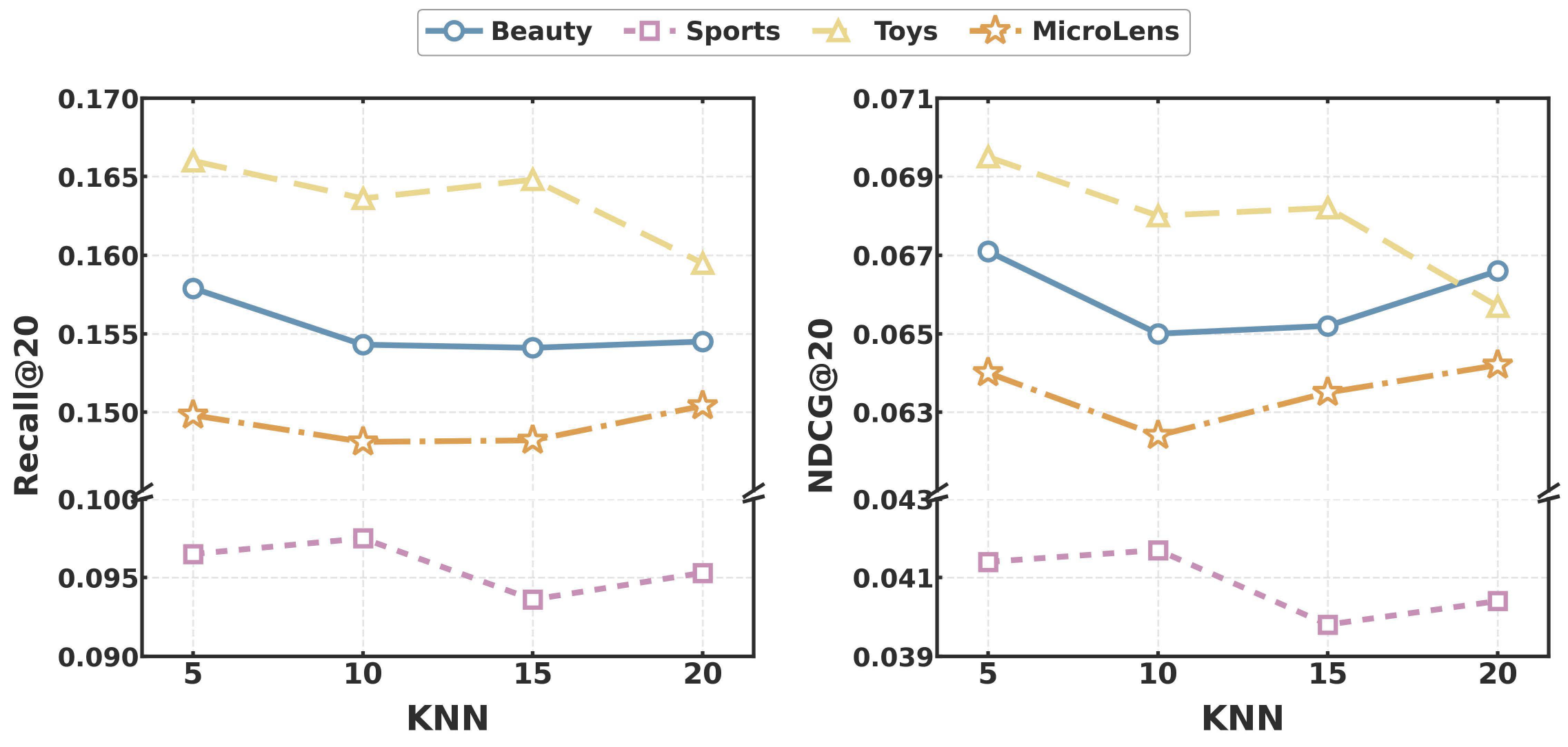}
   \caption{Effects of top-$K$ neighbors.}
   \Description{}
   \label{fig:3}
\end{figure}
\begin{figure}[t]
  \centering
   \includegraphics[width=\linewidth]{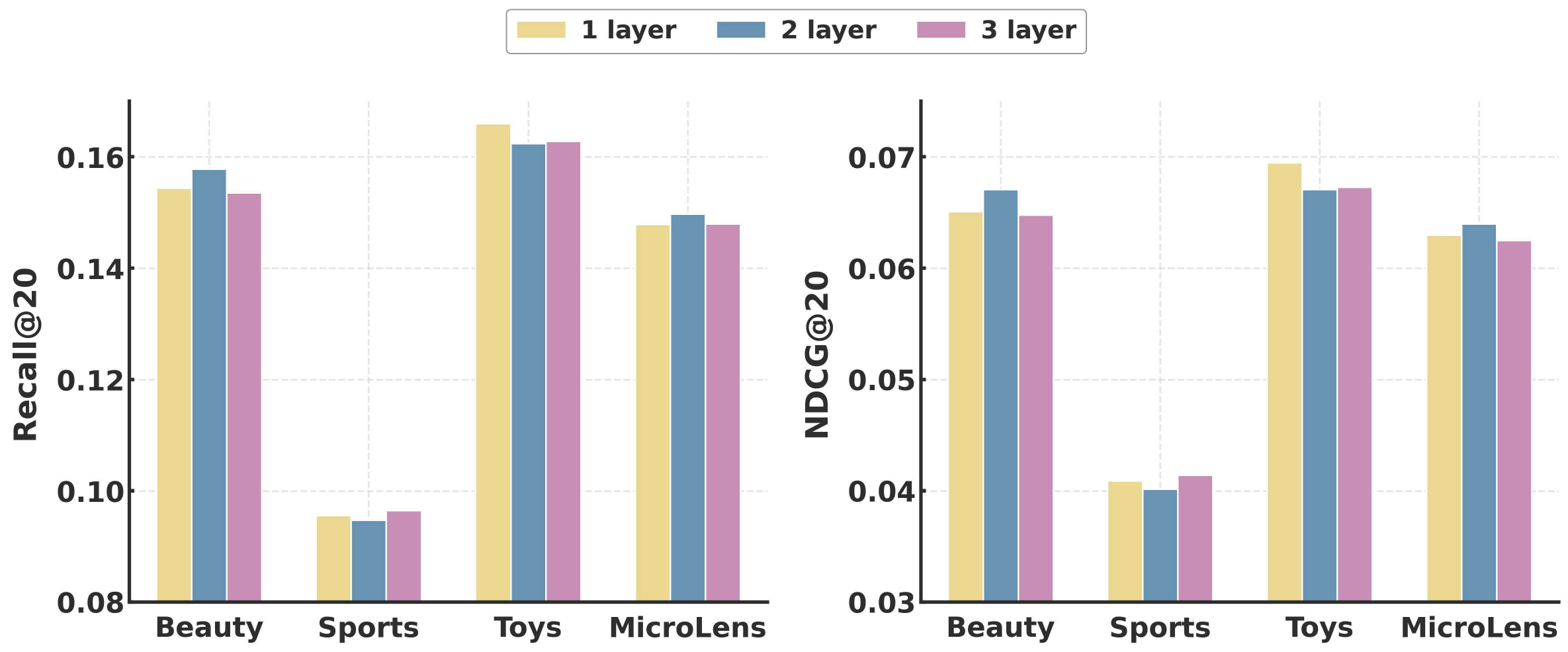}
   \caption{Effects of frequency-domain denoising layers.}
   \Description{}
   \label{fig:4}
\end{figure}
\subsection{Hyperparameter Sensitivity Analysis}

We investigate the sensitivity of two key hyperparameters.

\textit{Effect of Top-$K$ Neighbors.} Figure~\ref{fig:3} shows the impact of the number of top-$K$ neighbors ($K \in \{5, 10, 15, 20\}$) in item semantic graph construction. The optimal $K$ varies across datasets: Sports performs best at $K=10$, MicroLens at $K=20$, while Beauty and Toys achieve the best performance at $K=5$. We attribute this to differences in item scale and interaction density across datasets. For datasets with fewer items and sparse interactions (e.g., Beauty and Toys), a smaller $K$ better preserves the most relevant neighbors and avoids introducing noise. In contrast, datasets with larger item sets or denser interactions (e.g., Sports and MicroLens) benefit from a larger $K$, which captures richer semantic relationships. Despite these differences, performance varies slightly across $K$, demonstrating the robustness of the graph-based denoising module.

\textit{Effect of Frequency-domain Denoising Layers.} Figure~\ref{fig:4} illustrates the impact of the number of denoising layers ($L_f \in \{1, 2, 3\}$). The optimal $L_f$ varies across datasets: Beauty and MicroLens achieve the best performance at $L_f=2$, Toys at $L_f=1$, and Sports at $L_f=3$. As shown in the figure, performance generally improves when increasing $L_f$ from 1 to 2 on most datasets, but further increasing to 3 brings limited gains or slight degradation in some cases. This suggests a trade-off between noise suppression and information preservation. Datasets with higher interaction noise (e.g., Sports) benefit from deeper filtering, while cleaner datasets (e.g., Toys) may suffer from over-filtering when too many layers are applied.

\begin{figure}[t]
    \centering
    \includegraphics[width=\linewidth]{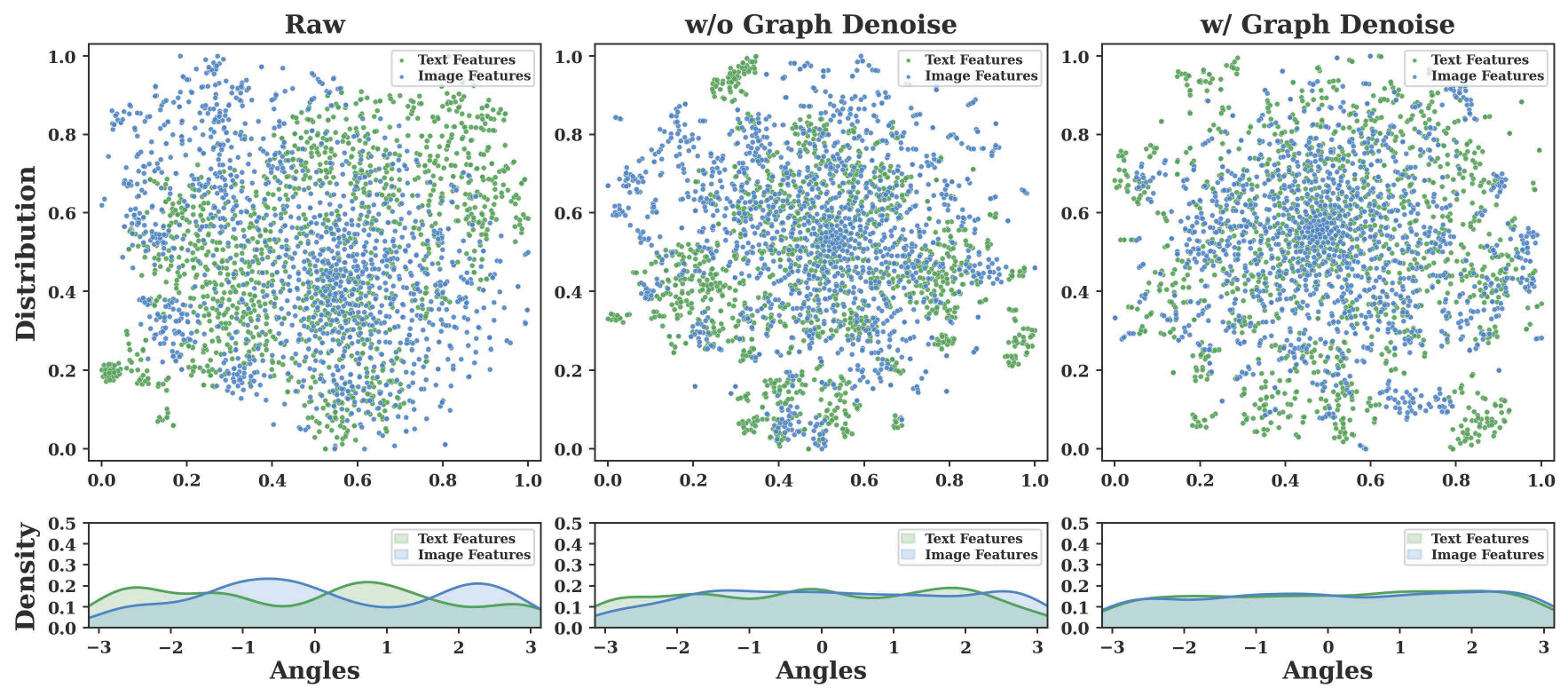}
    \Description{}
    \caption{Visualization of textual (green) and visual (blue) item feature distributions on the Beauty dataset under three settings. Top: 2D projections on the unit circle. Bottom: KDE of angular distributions. From left to right: \textbf{Raw}, \textbf{w/o Graph Denoise}, and \textbf{w/ Graph Denoise}.}
    \label{fig:5}
\end{figure}
\subsection{Visualization}
To intuitively demonstrate the effectiveness of the proposed graph-based feature denoising module, we visualize the multi-modal item feature distributions under three settings on the Beauty dataset:
(1) \textbf{Raw}: original features from frozen pre-trained encoders;
(2) \textbf{w/o Graph Denoise}: features from the full pipeline without graph-based denoising; and
(3) \textbf{w/ Graph Denoise}: features from the complete model.
For each setting, textual (green) and visual (blue) features are projected onto the unit circle, showing their 2D spatial distributions (top) and the kernel density estimation (KDE) of their angular distributions (bottom), as illustrated in Figure~\ref{fig:5}.

Two key observations can be drawn. \textbf{First}, in the \textbf{Raw} setting, the angular density curves of text and image features exhibit pronounced misalignment accompanied by multiple sharp, irregular peaks. This indicates a significant heterogeneity gap and severe anisotropy (i.e., dimensional collapse) in the pre-trained features. In the \textbf{w/o Graph Denoise} setting, where cross-modal contrastive learning is active, the distributions are partially smoothed. However, residual misalignment and irregular density fluctuations persist, suggesting that contrastive alignment alone struggles to overcome the intrinsic noise embedded in the raw representations. \textbf{Second}, with the integration of graph-based denoising (\textbf{w/ Graph Denoise}), the angular density curves of both modalities become markedly more uniform and tightly aligned, approaching a flat distribution across all angles. This demonstrates that our graph-based low-pass filtering effectively suppresses noisy semantic deviations at the source, yielding a highly isotropic feature space. Consequently, the synergistic combination of graph denoising and contrastive learning fundamentally mitigates the cross-modal distributional gap, providing high-quality, noise-free representations for downstream sequential recommendation.

\section{CONCLUSIONS}
In this paper, we address two complementary noise challenges in multi-modal sequential recommendation: redundant noise in pre-trained multi-modal features and interaction noise in user behavior sequences. We propose DDMSR, a dual-denoising framework that tackles both at their respective levels. At the feature level, a sparse item semantic graph is constructed to perform Laplacian smoothing as a 
low-pass filter, suppressing high-frequency noise in pre-trained representations while compensating for long-tail items via semantic propagation. At the sequence level, a learnable frequency-domain denoising module applies FFT/IFFT-based spectral modulation to adaptively attenuate noisy frequency components and preserve stable preference signals. A cross-modal contrastive objective further bridges the heterogeneity gap between textual and visual feature spaces for coherent multi-modal 
fusion. Extensive experiments on four public datasets demonstrate the effectiveness and superiority of DDMSR over state-of-the-art methods.


\bibliographystyle{ACM-Reference-Format}
\balance
\bibliography{content/references}

\appendix

\section{Appendix}
\subsection{Fourier Transform}

\noindent\textbf{Discrete Fourier Transform.}
Discrete Fourier Transform (DFT) is a fundamental tool in digital signal processing~\cite{dft:journals/tsmc/rabinergy78} that converts a sequence from the time domain to the frequency domain. Given a sequence $\{x_j\}$ with $j \in \{0, 1, \dots, n-1\}$, the DFT transforms it into the frequency domain according to:
\begin{equation}
    \tilde{x}_k = \sum_{j=0}^{n-1} x_j 
    \exp\!\left(-\frac{2\pi i}{n} jk\right), 
    \quad 0 \leq k \leq n-1,
\end{equation}
where $i$ is the imaginary unit, and $\tilde{x}_k$ denotes the spectrum of the sequence at frequency step $\omega_k = 2\pi k/n$. The DFT is a one-to-one transformation, and the original sequence can be recovered from its spectrum via the Inverse DFT (IDFT):
\begin{equation}
    x_j = \frac{1}{n} \sum_{k=0}^{n-1} \tilde{x}_k 
    \exp\!\left(\frac{2\pi i}{n} jk\right), 
    \quad 0 \leq j \leq n-1.
\end{equation}

\noindent\textbf{Fast Fourier Transform.}
The Fast Fourier Transform (FFT) is an efficient algorithm for computing the DFT, which reduces the time complexity from $\mathcal{O}(n^2)$ in the naive DFT to $\mathcal{O}(n \log n)$ by factorizing the DFT matrix into a product of sparse factors~\cite{fft:journals/siamrev/bailey93}. The inverse DFT can likewise be efficiently computed via the Inverse FFT (IFFT). Since FFT converts input signals into the frequency domain where periodic characteristics are more easily captured, it is widely adopted in sequential modeling to filter noise~\cite{fedformer:conf/icml/zhoumww0022, film:conf/nips/zhoumww0yy022,fmlp-rec:conf/www/zhouyzw22}.

In this work, we apply FFT and IFFT to the multi-modal item embedding sequence 
$\mathbf{H} \in \mathbb{R}^{n \times d}$ for sequence-level denoising. Specifically, FFT projects the sequence from the time domain into the frequency domain, where a learnable complex-valued filter modulates the spectrum to suppress noise-induced components; IFFT then reconstructs the purified sequence back to the time domain for downstream preference modeling.

\subsection{Details of Baselines} 
We compare our method with well-known SR baselines with two categories:

\noindent\textbf{ID-based Models:}
\begin{itemize}
    \item \textbf{SASRec}~\cite{sasrec:conf/icdm/kangm18} employs a 
    unidirectional self-attention mechanism to 
    capture sequential dependencies among items 
    for next-item prediction.
    
    \item \textbf{BERT4Rec}~\cite{bert4rec:conf/cikm/sunlwploj19} adopts 
    a bidirectional Transformer with a masked item 
    prediction task to learn richer sequential 
    representations.
    
    \item \textbf{FEARec}~\cite{fearec:conf/sigir/duyzqz0ls23} augments 
    self-attention with frequency-domain components 
    via a ramp structure to capture both low- and 
    high-frequency sequential signals.
    
    \item \textbf{BSARec}~\cite{bsarec:conf/aaai/shin0wp24} leverages 
    the Fourier transform to integrate frequency 
    information into self-attention, mitigating the 
    oversmoothing problem in sequential 
    recommendation. 
\end{itemize}

\noindent\textbf{Multi-modal Models:}
\begin{itemize}
    \item \textbf{UniSRec}~\cite{unisrec:conf/kdd/houmzldw22} learns 
    transferable universal sequence representations 
    from item textual descriptions via a 
    parametric whitening and MoE-enhanced adaptor, 
    enabling cross-domain recommendation.

    \item \textbf{MISSRec}~\cite{missrec:conf/mm/wangzwwllyzzx23} designs 
    an interest-aware Transformer encoder-decoder 
    that captures sequence-level multi-modal user 
    interests and adopts a dynamic fusion module 
    for user-adaptive item representations.

    \item \textbf{MP4SR}~\cite{mp4sr:journals/tors/zhangzzs25} proposes a 
    multi-modal pre-training framework that 
    leverages contrastive learning to capture 
    correlations among different modality sequences 
    of users and items for enhanced sequential 
    recommendation.

    \item \textbf{TedRec}~\cite{tedrec:conf/cikm/xutl0ww0cz24} transforms 
    textual and ID embeddings into the frequency 
    domain via Fourier transform, achieving 
    effective sequence-level text-ID semantic 
    fusion through element-wise multiplication.

    \item \textbf{HM4SR}~\cite{hm4sr:conf/www/zhang0s0025} proposes a 
    hierarchical time-aware Mixture-of-Experts 
    framework that integrates explicit temporal 
    information with multi-modal features to model 
    dynamic user interests.

    \item \textbf{DMMD4SR}~\cite{dmmd4sr:conf/mm/luy25} introduces 
    a diffusion model-based multi-level denoising 
    framework that progressively mitigates domain 
    shift noise and interest-agnostic noise in 
    multi-modal representations for sequential 
    recommendation.
\end{itemize}
For a fair comparison, we train UniSRec, MP4SR and MISSRec from scratch without pre-training on additional datasets.

\begin{table}[t]
    \centering
    \caption{Efficiency comparison with DMMD4SR. \textit{Time/E} (s) denotes the average training time per epoch, and \textit{MU} (GB) indicates GPU memory usage.}
    \label{tab:efficiency}
    \resizebox{0.48\textwidth}{!}{
        \begin{tabular}{c|c|c|c|c|c}
            \toprule
            Datasets & Model & Recall@20 & NDCG@20 & Time/E & MU \\        
            \midrule
            \multirow{2}{*}{Beauty} 
            & DMMD4SR & 0.1129 & 0.0466 & 42s & 18.13G \\
            & DDMSR & 0.1579 & 0.0671 & 14s & 9.32G \\
            \midrule
            \multirow{2}{*}{Sports} 
            & DMMD4SR & 0.0672 & 0.0262 & 62s & 18.48G \\
            & DDMSR & 0.0975 & 0.0417 & 26s & 12.36G \\
            \midrule
            \multirow{2}{*}{Toys} 
            & DMMD4SR & 0.1159 & 0.0488 & 36s & 18.13G \\
            & DDMSR & 0.1660 & 0.0695 & 11s & 8.56G \\
            \midrule
            \multirow{2}{*}{MicroLens} 
            & DMMD4SR & 0.1216 & 0.0490 & 136s & 18.44G \\
            & DDMSR & 0.1497 & 0.0637 & 53s & 11.15G \\
            \bottomrule
        \end{tabular}
    }
\end{table}
\subsection{Efficiency Analysis}
\label{sec:efficiency}

In this section, we comprehensively compare the training efficiency of our proposed DDMSR against the generative diffusion-based baseline, DMMD4SR, from both theoretical complexity and empirical perspectives.

Let $B$ denote the batch size, $L$ the maximum sequence length, $d$ the hidden feature dimension, $N$ the item vocabulary size, and $k$ the number of retained neighbors in the sparse similarity graph.

\textbf{Theoretical Training Complexity.} 
The denoising module of DMMD4SR heavily relies on multiple generative diffusion networks. During the training phase, for each batch, DMMD4SR not only computes the reconstruction loss but also executes a $T$-step autoregressive reverse sampling across $M$ independent diffusion networks (e.g., text, image, and sequence SDNets) to obtain denoised representations for subsequent sequence encoding. Since the single-step prediction of each network is dominated by MLPs with a complexity of $\mathcal{O}(B \cdot L \cdot d^2)$, the total time complexity per training batch reaches:
\begin{equation}
    \mathcal{O}_{\text{DMMD4SR}} = \mathcal{O}(M \cdot T \cdot B \cdot L \cdot d^2)
\end{equation}
Furthermore, storing the multi-step computational graphs and gradients for all modality experts incurs an extremely high memory footprint.

In contrast, DDMSR introduces a highly efficient discriminative spatial-frequency denoising paradigm. For the Graph Feature Denoiser, graph aggregation over the entire item vocabulary is performed via sparse matrix multiplication, incurring a per-forward cost of $\mathcal{O}(N \cdot k \cdot d)$. For the sequence Frequency Filter, high-frequency noise is suppressed in a single one-shot operation: across $F$ stacked filter layers, the Fast Fourier Transform (FFT) and its inverse contribute $\mathcal{O}(F \cdot B \cdot d \cdot L\log L)$, while the sequential gating mechanism contributes $\mathcal{O}(F \cdot B \cdot L \cdot d^2)$. The combined per-batch training complexity of DDMSR is therefore:
\begin{equation}
    \mathcal{O}_{\text{DDMSR}} = \mathcal{O}\!\left(N \cdot k \cdot d \; +\; F \cdot B \cdot L \cdot d \cdot \left(\log L + d\right) \right).
    \label{eq:ddmsr_complexity}
\end{equation}
It is worth noting that although Eq.~\eqref{eq:ddmsr_complexity} contains the term $\mathcal{O}(N \cdot k \cdot d)$, this operation amounts to a single, highly optimized sparse aggregation step whose practical wall-clock overhead is negligible compared to the repeated multi-step MLP evaluations in diffusion-based denoising. The training complexity of DDMSR can therefore be effectively regarded as $\mathcal{O}\!\left(F \cdot B \cdot L \cdot d \cdot \left(\log L + d\right)\right)$. Moreover, during online inference, the item-level graph aggregation can be pre-computed entirely offline, eliminating the $\mathcal{O}(N \cdot k \cdot d)$ term from the online critical path and reducing its effective inference-time cost to $\mathcal{O}(1)$.

\textbf{Empirical Efficiency.} 
The theoretical advantages are strongly validated by the empirical training overhead presented in Table~\ref{tab:efficiency}. Across all evaluated datasets, DDMSR consistently achieves a 2.5$\times$ to 3$\times$ speedup in training time per epoch (Time/e) compared to DMMD4SR (e.g., reducing the training time from 42s to 14s on the Beauty dataset). Additionally, by completely circumventing the cumbersome multi-step diffusion graphs, DDMSR reduces GPU memory usage (MU) by roughly 30\% to 50\%. Most importantly, while substantially reducing both computational and memory costs, DDMSR still yields significant performance improvements in Recall@20 and NDCG@20. This firmly demonstrates the superiority, scalability, and practical value of our proposed method in real-world recommender systems.

\begin{figure}[t]
  \centering
   \includegraphics[width=\linewidth]{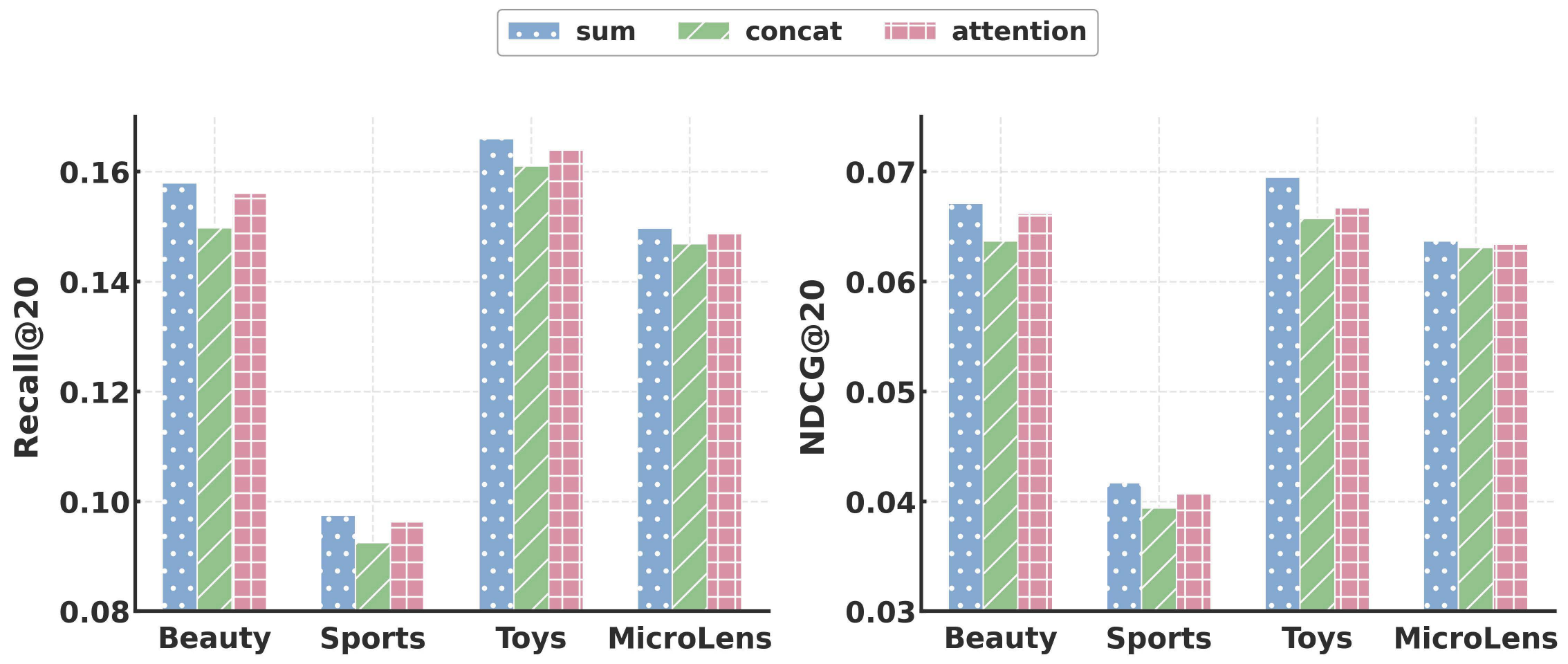}
   \caption{Effects of the multi-modal fusion type.}
   \Description{}
   \label{fig:6}
\end{figure}
\begin{figure}[t]
  \centering
   \includegraphics[width=\linewidth]{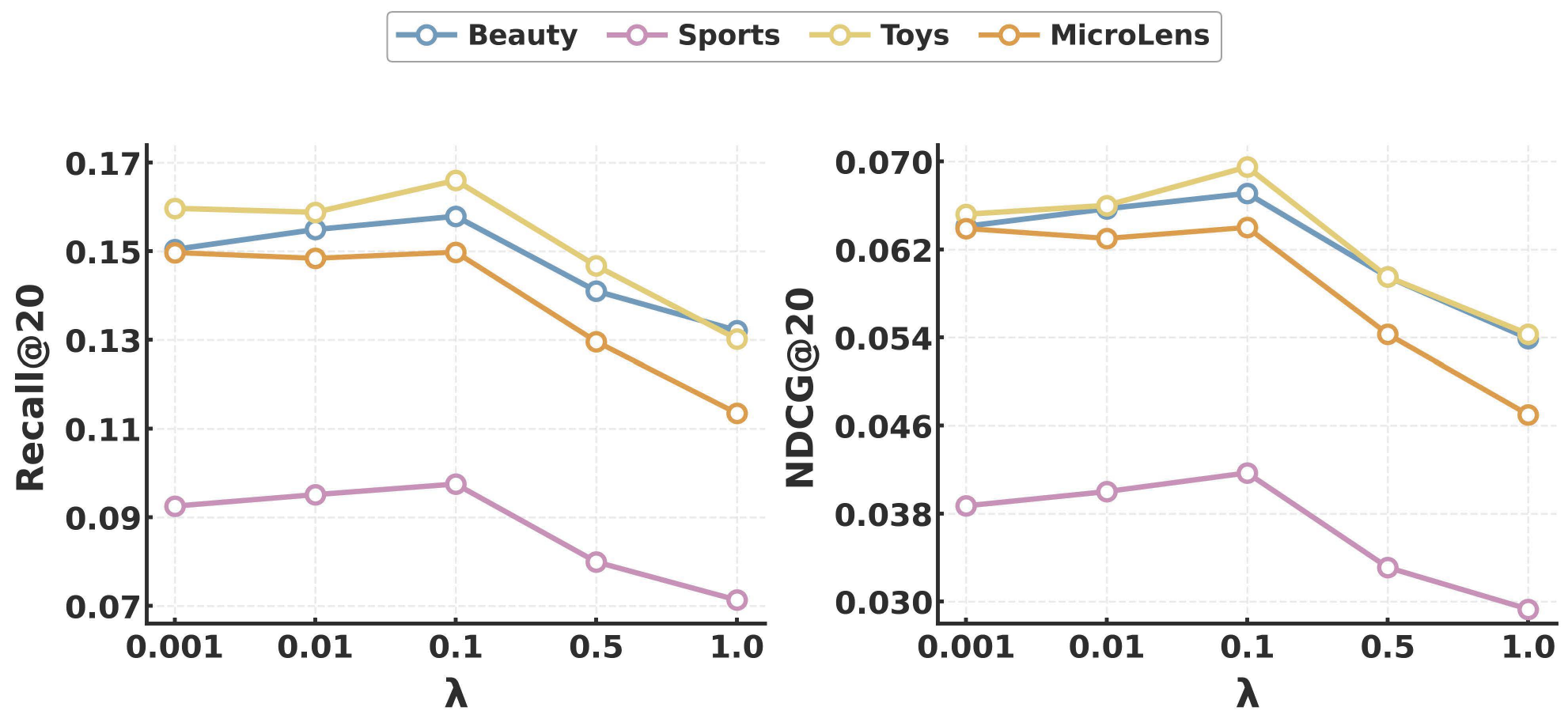}
   \caption{Effects of the contrastive learning weight $\lambda$.}
   \Description{}
   \label{fig:7}
\end{figure}
\subsection{More Hyperparameter Sensitivity Analysis}
\label{app:hyperparameter}

Due to space limitations in the main text, we present further hyperparameter sensitivity analyses in this section. Specifically, we investigate the impact of different multi-modal fusion strategies and the contrastive learning weight $\lambda$ on the overall recommendation performance.

\noindent\textbf{Impact of Multi-modal Fusion Strategies.} 
To determine the optimal operation for integrating textual and visual features, we evaluate three commonly used fusion functions for $\text{Fusion}(\cdot)$ in Eq.9: element-wise addition (\textit{add}), concatenation followed by a linear projection (\textit{concat}), and an attention mechanism (\textit{attention}). As illustrated in Figure~\ref{fig:6}, element-wise addition consistently outperforms the other two variants across all datasets in terms of both Recall@20 and NDCG@20. We attribute this to the fact that element-wise addition provides a symmetric and parameter-free mechanism to superimpose multi-modal signals. It effectively aggregates complementary semantic information without altering the original feature space dimensionality. In contrast, the \textit{concat} operation introduces extra learnable parameters that may lead to overfitting, particularly in sparse recommendation scenarios. Similarly, the \textit{attention} mechanism can be overly complex and computationally unstable for fusing already aligned modality embeddings. Consequently, we adopt element-wise addition as the default fusion strategy in our framework.

\noindent\textbf{Impact of Contrastive Learning Weight $\lambda$.} 
The hyperparameter $\lambda$ controls the magnitude of the multi-modal contrastive alignment objective, which balances the auxiliary cross-modal regularization with the primary sequential recommendation task. We search for the optimal $\lambda$ within the range of $\{0.001, 0.01, 0.1, 0.5, 1.0\}$. As depicted in Figure~\ref{fig:7}, the performance experiences a steady improvement as $\lambda$ increases from $0.001$, reaching a consistent peak at $\lambda = 0.1$. However, continuously increasing $\lambda$ beyond $0.1$ (e.g., to $0.5$ or $1.0$) leads to a sharp performance drop. This trend aligns with our intuition: a moderate contrastive weight effectively bridges the heterogeneity gap between modalities and enforces semantic consistency. Conversely, an excessively large $\lambda$ causes the model to over-focus on the alignment task, thereby dominating the gradient updates and inducing negative transfer to the main temporal preference modeling. Hence, we set $\lambda = 0.1$ for all main experiments.

\subsection{Long-Tail Analysis}
\begin{figure}[h]
  \centering
   \includegraphics[width=\linewidth]{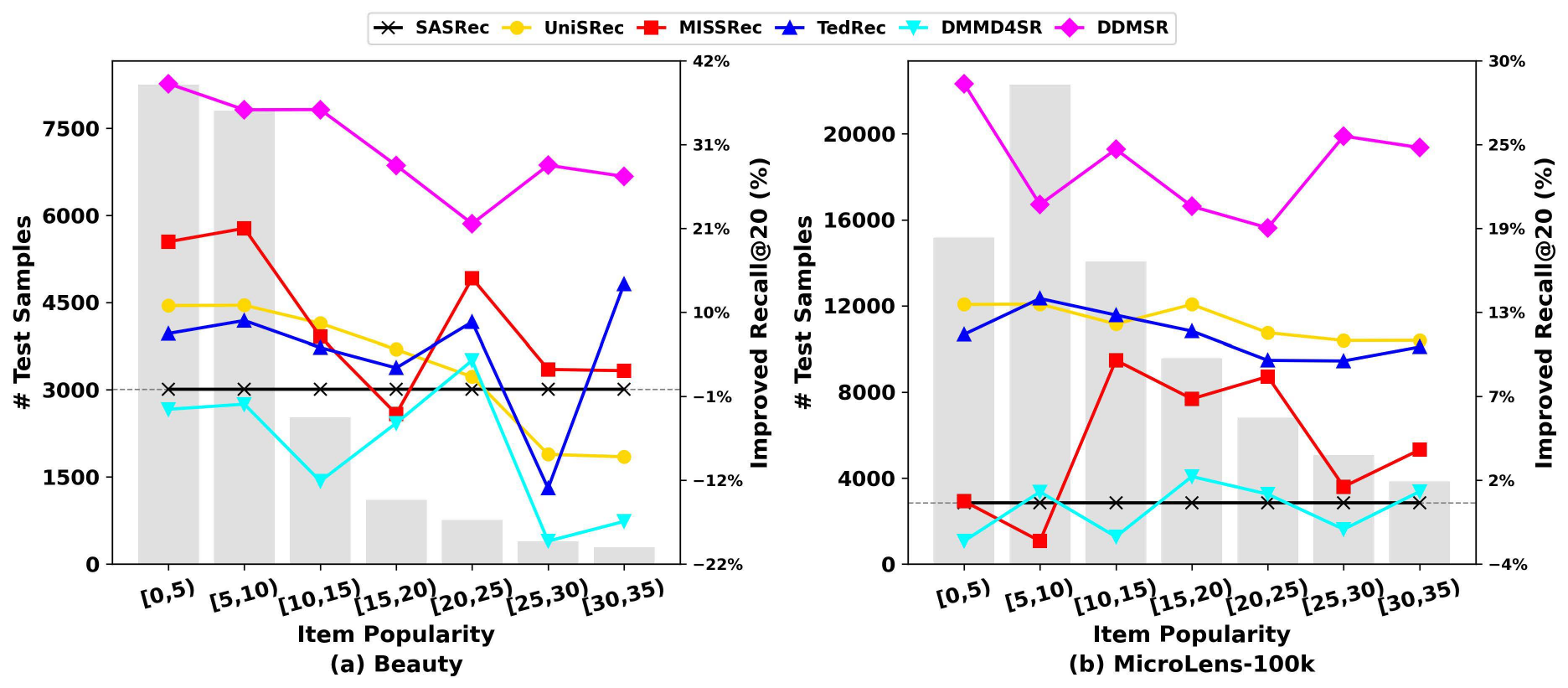}
   \caption{Performance improvement across different item popularity (Long-tail) groups.}
   \Description{}
   \label{fig8}
\end{figure}

Pure ID-based models (e.g., SASRec) inherently suffer from representation collapse on data-sparse items. In Fig.~\ref{fig8}, we establish the pure ID model SASRec as the baseline to evaluate relative performance improvements. While other multi-modal models exhibit moderate robustness on infrequent items, DDMSR decisively dominates. It achieves massive, unprecedented performance gains on the least frequent items (e.g., the $[0, 5)$ and $[5, 10)$ bins). This directly confirms that our graph-based semantic denoising, coupled with cross-modal alignment, successfully leverages neighborhood aggregation and modality complementarity to enrich high-quality multi-modal representations for long-tail items where interactions are extremely scarce.

\begin{table}[h]
    \centering
    \caption{Robustness under Noise (Recall@20)}
    \label{tab:robustness}
    \resizebox{0.48\textwidth}{!}{
        \begin{tabular}{c|c|c|cc|cc|cc}
            \toprule
            \multirow{2}{*}{Dataset} & \multirow{2}{*}{Method} & Clean & \multicolumn{2}{c|}{$n_r=0.1$} & \multicolumn{2}{c|}{$n_r=0.3$} & \multicolumn{2}{c}{$n_r=0.5$} \\
            \cmidrule(lr){4-5} \cmidrule(lr){6-7} \cmidrule(lr){8-9}
            & & R@20 & R@20 & Drop & R@20 & Drop & R@20 & Drop \\
            \midrule
            \multirow{3}{*}{Beauty}
            & UniSRec       & 0.1265 & 0.1047 & -17.2\% & 0.0961 & -24.0\% & 0.0677 & -46.5\% \\
            & DMMD4SR       & 0.1109 & 0.0958 & -13.6\% & 0.0902 & -18.7\% & 0.0640 & -42.3\% \\
            & DDMSR        & 0.1579 & 0.1405 & \textbf{-11.0\%} & 0.1337 & \textbf{-15.3\%} & 0.1039 & \textbf{-34.2\%} \\
            \midrule
            \multirow{3}{*}{MicroLens}
            & UniSRec       & 0.1374 & 0.1139 & -17.1\% & 0.1079 & -21.4\% & 0.0756 & -45.0\% \\
            & DMMD4SR       & 0.1219 & 0.1012 & -17.0\% & 0.0962 & -21.1\% & 0.0661 & -45.8\% \\
            & DDMSR        & 0.1499 & 0.1337 & \textbf{-10.8\%} & 0.1302 & \textbf{-13.1\%} & 0.1011 & \textbf{-32.6\%} \\
            \bottomrule
        \end{tabular}
    }
\end{table}
\subsection{Noise Robustness} 
As clarified, our feature denoising targets semantic artifacts, not random numerical fluctuations. Injecting synthetic random noise corrupts the representation manifold, causing graph aggregation to inadvertently amplify unstructured interference. Furthermore, realistically simulating multi-modal semantic noise (e.g., altering backgrounds) is complex and introduces confounders. Therefore, to strictly evaluate noise robustness, we conducted sequence-level perturbations to simulate the common noise in recommendation systems. Specifically, we replaced a proportion ($n_r \in \{0.1, 0.3, 0.5\}$) of actual user interactions with random items to mimic severe accidental clicks. While all models degrade due to sequence corruption, DDMSR exhibits the strongest resilience. As shown in Table~\ref{tab:robustness}, under extreme 50\% noise on MicroLens, DDMSR's Recall@20 drops by only 32.6\%, significantly outperforming DMMD4SR (45.8\% drop) and UniSRec (45.0\% drop). This rigorously proves that our frequency filter actively attenuates anomalous signals rather than acting as a standard transformation.

\end{document}